\documentclass[%
reprint,twocolumn,
prx,
manuscript=article
]{revtex4-2}

\usepackage{graphicx}
\usepackage{bm}
\usepackage{amsmath}
\usepackage{amssymb}
\usepackage{appendix}
\usepackage{comment}
\usepackage{relsize}
\usepackage[bottom]{footmisc}
\usepackage[dvipsnames]{xcolor}
\usepackage[section]{placeins}
\usepackage{layouts}
\usepackage{bbold}
\usepackage{enumitem}
\usepackage{subcaption}
\definecolor{darkblue}{rgb}{0,0,.65}
\definecolor{darkgreen}{rgb}{0.28,0.41,0.19}

\usepackage{braket}
\usepackage[version=3]{mhchem}
\usepackage{tikz}
\usepackage{caption}
\usetikzlibrary{shapes, arrows, chains, shapes.multipart}
\usepackage[normalem]{ulem}
\usepackage{bbm}

\newcommand{\tr}{\mathrm{Tr}}

\newcommand{\up}{\uparrow}
\newcommand{\dw}{\downarrow}
\newcommand{\fqi}{F_\mathrm{QI}}

\begin{document}

\title{Unveiling Intrinsic Many-Body Complexity by Compressing Single-Body Triviality}

\author{Ke Liao}
\email{ke.liao.whu@gmail.com}
\affiliation{Faculty of Physics, Arnold Sommerfeld Centre for Theoretical Physics (ASC),\\Ludwig-Maximilians-Universit{\"a}t M{\"u}nchen, Theresienstr.~37, 80333 M{\"u}nchen, Germany}
\affiliation{Munich Center for Quantum Science and Technology (MCQST), Schellingstrasse 4, 80799 M{\"u}nchen, Germany}

\author{Lexin Ding}
\affiliation{Faculty of Physics, Arnold Sommerfeld Centre for Theoretical Physics (ASC),\\Ludwig-Maximilians-Universit{\"a}t M{\"u}nchen, Theresienstr.~37, 80333 M{\"u}nchen, Germany}
\affiliation{Munich Center for Quantum Science and Technology (MCQST), Schellingstrasse 4, 80799 M{\"u}nchen, Germany}

\author{Christian Schilling}
\email{c.schilling@physik.uni-muenchen.de}
\affiliation{Faculty of Physics, Arnold Sommerfeld Centre for Theoretical Physics (ASC),\\Ludwig-Maximilians-Universit{\"a}t M{\"u}nchen, Theresienstr.~37, 80333 M{\"u}nchen, Germany}
\affiliation{Munich Center for Quantum Science and Technology (MCQST), Schellingstrasse 4, 80799 M{\"u}nchen, Germany}

\begin{abstract}
   The simultaneous treatment of static and dynamic correlations in strongly-correlated electron
  systems is a critical challenge.  In particular, finding a universal scheme for identifying a single-particle orbital basis that minimizes the representational complexity of the many-body wavefunction is a formidable and longstanding problem. As a contribution towards its solution, we show that the total orbital correlation actually reveals and quantifies the intrinsic complexity of the wavefunction,
  once it is minimized via orbital rotations.  To demonstrate the power of this concept in practice,
  an iterative scheme is proposed to optimize the orbitals by minimizing the total orbital correlation calculated by
  the tailored coupled cluster singles and doubles (TCCSD) ansatz.
 The optimized orbitals enable the limited TCCSD ansatz
  to capture more non-trivial information of the many-body wavefunction,
  indicated by the improved wavefunction and energy.
  An initial application of this scheme
  shows great improvement of TCCSD in predicting the
  singlet ground state potential energy curves of the
  strongly correlated  C$_{\rm 2}$ and Cr$_{\rm 2}$ molecule.
\end{abstract}

\maketitle

\begin{figure}[ht]
  \centering
  \includegraphics[scale=1.3]{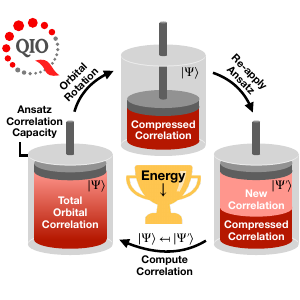}
  \caption{TOC graphic}
\end{figure}

The \emph{ab-initio} simulation of strongly correlated electron systems is a
central challenge in quantum chemistry~\cite{sharma2014a,kurashige2013b} and
materials science~\cite{Booth2013,bogdanov2022,cui2022}. More specifically, this
problem is complicated by the intricate interplay between the so-called static
and dynamic electron-electron correlations. The former exhibits itself in the
presence of many significant many-body configurations, or in the case of matrix
product states (MPS) the very high bond dimensions, needed to even qualitatively
describe the wavefunction.
%One of the common causes is degeneracies related to some underlying symmetries.
The latter includes the remaining correlation effects, such as those related to
the short-range Kato cusp condition~\cite{Kato1957}, long-range
electron-electron screening
effects~\cite{Shepherd2013a,Ochi2017,masios2023a,neufeld2023a}, van der Waals
interactions~\cite{drummond2007a,tkatchenko2012,cioslowski2023}, etc. Towards
solving this problem, recent years have witnessed promising developments of a
plethora of methods, including tensor network
theories~\cite{schollwock2005a,chan2011a,baiardi2020}, quantum Monte Carlo
methods~\cite{Booth2009a,cleland2010,limanni2016,shi2021,lee2022}, quantum
chemical theories, such as specific perturbation
theories~\cite{andersson1992,angeli2001,limanni2014,gagliardi2017}, coupled
cluster (CC)~\cite{Bartlett2007-cm,Kats2013d,kats2014a,gruber2018a,kats2019} and
functional theories~\cite{yang2000,mazziotti2006a,mazziotti2008,schilling2021}.
Each of these methods covers certain pieces of the strong correlation puzzle.
Ongoing efforts are focussed on utilizing combinations of them to reach larger
areas of the correlation landscape~\cite{neuscamman2010, Luo2018, vitale2020,
liao2021b, schraivogel2021a, baiardi2022, liao2023}, while maintaining a
delicate balance between accuracy and efficiency. In face of the complexity of
the total electron correlation problem, it is surprising that the pivotal role
of the underlying single-particle orbital basis is neither fully appreciated nor
conclusively understood yet.

For systems dominated by dynamic correlations, the choice of the orbitals
influences more the efficiency than the accuracy: Typically, orbitals from
cost-effective mean-field theories like Hartree-Fock (HF) and density functional
theory (DFT) are used to construct correlated many-body wavefunction ans\"atze;
Natural orbitals, introduced by Löwdin~\cite{lowdin1955a}, are also commonly
used for a more compact Slater determinant expansion in configuration
interaction (CI), CC~\cite{taube2005a,taube2008b,Gruneis2011b,liao2019a}, and
explicitly correlated
methods~\cite{kong2012b,ma2017,pavosevic2017,kallay2023,dobrautz2023b};
Localized orbitals~\cite{li2009,rolik2013} are often used to achieve linear
scaling with system size, exploiting the entanglement area law in gapped
systems~\cite{srednicki1993,hastings2007,eisert2010a}.

In systems with strong static correlations, such as systems containing
transition metal elements with partially filled $d$-shells, the choice of
orbitals can significantly impact not only the efficiency, but, more
importantly, the accuracy. It is well recognized that the choice of the
single-particle orbitals affects the representational complexity of the
many-body wavefunction, e.g.~in the context of the density matrix
renormalization group (DMRG)~\cite{krumnow2016} and full configuration
interaction quantum Monte Carlo (FCIQMC)~\cite{limanni2020}. A badly chosen
orbital basis can lead to an artificially difficult computational problem. For
example, HF orbitals are normally too delocalized to capture strongly correlated
physics, such as strong on-site Coulomb interactions and spin fluctuations. An
important advancement was the introduction of the complete active space (CAS)
self-consistent field (CASSCF) method~\cite{roos1980a,siegbahn1981}, where both
the orbitals and the CI coefficients in the active space are optimized
simultaneously. Another popular choice are the natural orbitals (NO) from
computationally efficient methods~\cite{dobrautz2023b,larsson2022,vitale2020}.
For a comprehensive review of the active space orbital construction and
selection methods, we refer to Ref.~\cite{toth2020} and the references therein
\footnote{The selection of the active space in general is a very challenging
task on its own, here we focus on the optimization of the orbitals within a
given active space size.}.

As a key motivation for our work, we recall that none of the aforementioned
orbitals are optimal for the total correlation problem as they are optimized for
either static or dynamic correlations, but not for both. Some recognition of the
importance of optimizing orbitals by taking into account both types of
correlations exists at the level of single reference
methods~\cite{lee2018a,kats2014a}. Yet, a general scheme that can be easily
extended to other higher-level theories is still missing. The main reason for
this is the lack of a concise tool that allows one to quantify the
\emph{intrinsic complexity} of the many-body wavefunction universally, i.e.,
independently of the underlying ans\"atze. Inspired by previous work using
effective quantum information concepts in quantum
chemistry~\cite{boguslawski2012,boguslawski2015,stein2016,stein2017,ding2020,ding2021},
it will be the accomplishment of our work to provide exactly this missing tool.
To be more specific, we will establish the \emph{total orbital correlation},
defined in \eqref{equ:I}, as a quantitative means to link the single-particle
basis and the many-body wavefunction's representational complexity, to be
explained below in the next paragraphs. In particular, we will demonstrate
that through minimizing the total orbital correlation one can systematically
reduce the representational complexity, and hence reveal the intrinsic
complexity of the many-body wavefunction. In practice, this leads to an improved
accuracy of approximate ans\"atze like TCCSD for strongly correlated systems, as
we will show in the exemplary study of the C$_{\rm 2}$ and Cr$_{\rm 2}$
molecule.

To introduce and establish the total orbital correlation as a quantitative means
for describing the representational complexity of the many-body wavefunction, we
first recall some basic quantum information concepts, along with an illustrative
example. Given an orbital basis $\mathcal{B}$, and a particle number and spin
conserving ground state wavefunction $\ket{\Psi}$, we can always perform the
following Schmidt decomposition with respect to the bipartition between an
orbital $i$ and the rest of the orbitals, $ \ket{\Psi} =
\sum_{k={0,\up,\dw,\up\dw}} \sqrt{\lambda^{(i)}_k}\ket{k}_i \otimes
\ket{\psi_k}_{\backslash\{i\}}$, where $\lambda^{(i)}_k$ are the eigenvalues of
the reduced density matrix $\rho_i$ of orbital $i$ defined as $\rho_i =
\mathrm{Tr}_{\backslash\{i\}}[\ket{\Psi}\bra{\Psi}]$. Accordingly, the entropy
of orbital $i$ follows as
\begin{equation}
  S(\rho_i) \equiv -\tr [\rho_i \log \rho_i]= -\sum_{k={0,\up,\dw,\up\dw}} \lambda^{(i)}_k \log \lambda^{(i)}_k, \label{equ:orb_ent}
\end{equation}
which quantifies for pure states precisely the entanglement between orbital $i$
and the rest of the orbitals, or equivalently (up to a factor $1/2$) the entire
correlation including both quantum and classical
contributions\cite{henderson2001,groisman2005}. It is worth noticing here that
the eigenvalues $\lambda_k^{(i)}$ change as $\mathcal{B}$ is varied. This can
easily be seen when they are expressed as functions of the one- and two-particle
reduced density matrix (1-RDM and
2-RDM)~\cite{boguslawski2015,ding2021,ding2023b}. Moreover, the \emph{total}
orbital correlation quantifies through the quantum relative entropy
$S(\rho||\sigma)\equiv \tr [\rho (\log(\rho)-\log(\sigma))]$ the deviation of
the quantum state $\rho$ (including the case of mixed states $\rho \neq
|\Psi\rangle \langle \Psi |$) from the manifold of states with zero correlation
between various orbitals in $\mathcal{B}$, i.e.
\begin{eqnarray}\label{equ:I} I_{\mathrm{tot}}^{(\mathcal{B})}(\rho) &:=&
\min_{\sigma \equiv \sigma_1  \otimes \ldots \otimes \sigma_d}
S\big(\rho\big|\big| \sigma\big) \nonumber \\
&=& \sum_{i=1}^{M} S(\rho_i)-S(\rho).
\end{eqnarray}
In the last line, we used the well-know fact~\cite{modi2010} that the minimum in
the first line is attained for the uncorrelated state given by the product of
the single orbital reduced density matrices $\rho_i$ of $\rho$.  In that sense
$I_{\mathrm{tot}}^{(\mathcal{B})}(\rho)$ quantifies the correlation of various
$M$ orbitals in $\mathcal{B}$ collectively.

In order to illustrate how the total orbital correlation directly reflects the
multiconfigurational character of a wavefunction $\ket{\Psi}$, we first consider
a two-electron singlet state in two orbitals. In this case, the Shannon entropy
of the CI coefficients of the many-body wavefunction expanded in the four
configurations $\{\ket{0,\uparrow\downarrow}, \ket{\uparrow,\downarrow},
\ket{\downarrow,\uparrow},\ket{\uparrow\downarrow,0}\}$ is defined as
$H(\{|c_i|^2\})=-\sum_{i=1}^4|c_i|^2\log(|c_i|^2)$ which actually equals (up to
a prefactor) the total orbital correlation \eqref{equ:I}. In particular, when
the total orbital correlation is 0, the state $\ket{\Psi}$ is a single Slater
determinant, whereas when it is maximal, $\ket{\Psi}$ is also maximally
multiconfigurational. Remarkably, the total orbital correlation of $\ket{\Psi}$
can vary drastically when the orbital basis changes.
\begin{figure}[htb]
  \centering
  \includegraphics[scale=0.6]{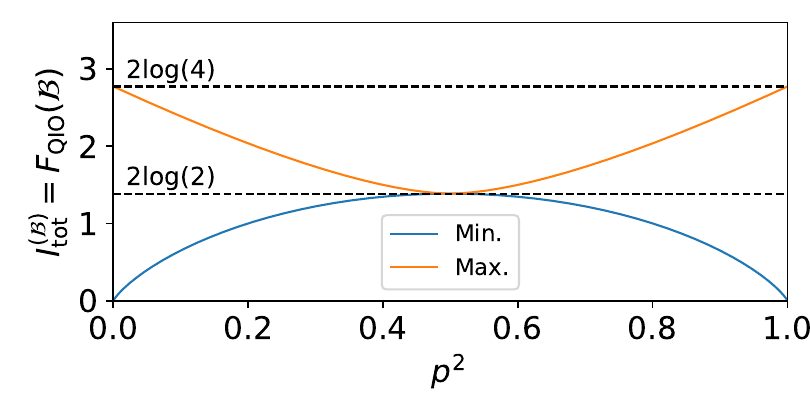}
  \caption{Minimal and maximal total orbital correlation, Eq.~\eqref{equ:I},
  over all real orbital bases $\mathcal{B}$ for a two electron singlet state
  $\ket{\Psi} = p \ket{\uparrow\downarrow,0} +
  \sqrt{1-p^2}\ket{0,\uparrow\downarrow}$ against $p^2$.}
  \label{fig:2el}
\end{figure}
One can verify that it attains its minimal value for the natural orbitals of
$\ket{\Psi}$ (which are in general not the minimizer), where $\ket{\Psi}$ can be
expressed as a zero-seniority state $\ket{\Psi} = p \ket{\uparrow\downarrow,0} +
\sqrt{1-p^2}\ket{0,\uparrow\downarrow}$, where we use $p$ to denote the CI
coefficient of the first configuration in the natural orbital basis.
%In this basis, the redundant multireference character of $\ket{\Psi}$ due to a
%poorly chosen orbital basis is optimally transformed away.
We therefore define the \emph{single-body triviality} as any redundant total
orbital correlation beyond the minimal total orbital correlation, as shown as
the blue curve in Fig.~\ref{fig:2el}, and the \emph{representational complexity}
as the value of the total orbital correlation in the current orbital basis. When
$p^2$ is closed to 0 or 1, $\ket{\Psi}$ is of single reference character, and
correspondingly the difference of the total orbital correlation between the
``best'' and ``worst'' choice of orbital basis is radical. When $p^2=1/2$, the
state is equally and maximally multiconfigurational in every orbital basis,
which means there is no single-body triviality in the state. For more details on
this example, we refer to Appendix \ref{app:2el}.
%no advantage can be gained by orbital rotations.

Based on the analytic insights above, and the precise meaning of the total
orbital correlation, we propose the following cost function 
\begin{equation}
  F_\mathrm{QIO}(\mathcal{B}) \equiv I_{\mathrm{tot}}^{(\mathcal{B})}(|\Psi\rangle\langle\Psi|)= \sum_{i=1}^{M} S(\rho_i),
  \label{equ:qio_cost}
\end{equation}
where $M$ is the total number of spatial orbitals. The minimization of
\eqref{equ:qio_cost} leads to the quantum information orbitals (QIO).
The cost function is 0 when the wavefunction is a single Slater determinant, and
achieves the maximal value of $M\log 4$ when the wavefunction is maximally
multiconfigurational.

\begin{figure}[ht]
\centering
\includegraphics[scale=0.25]{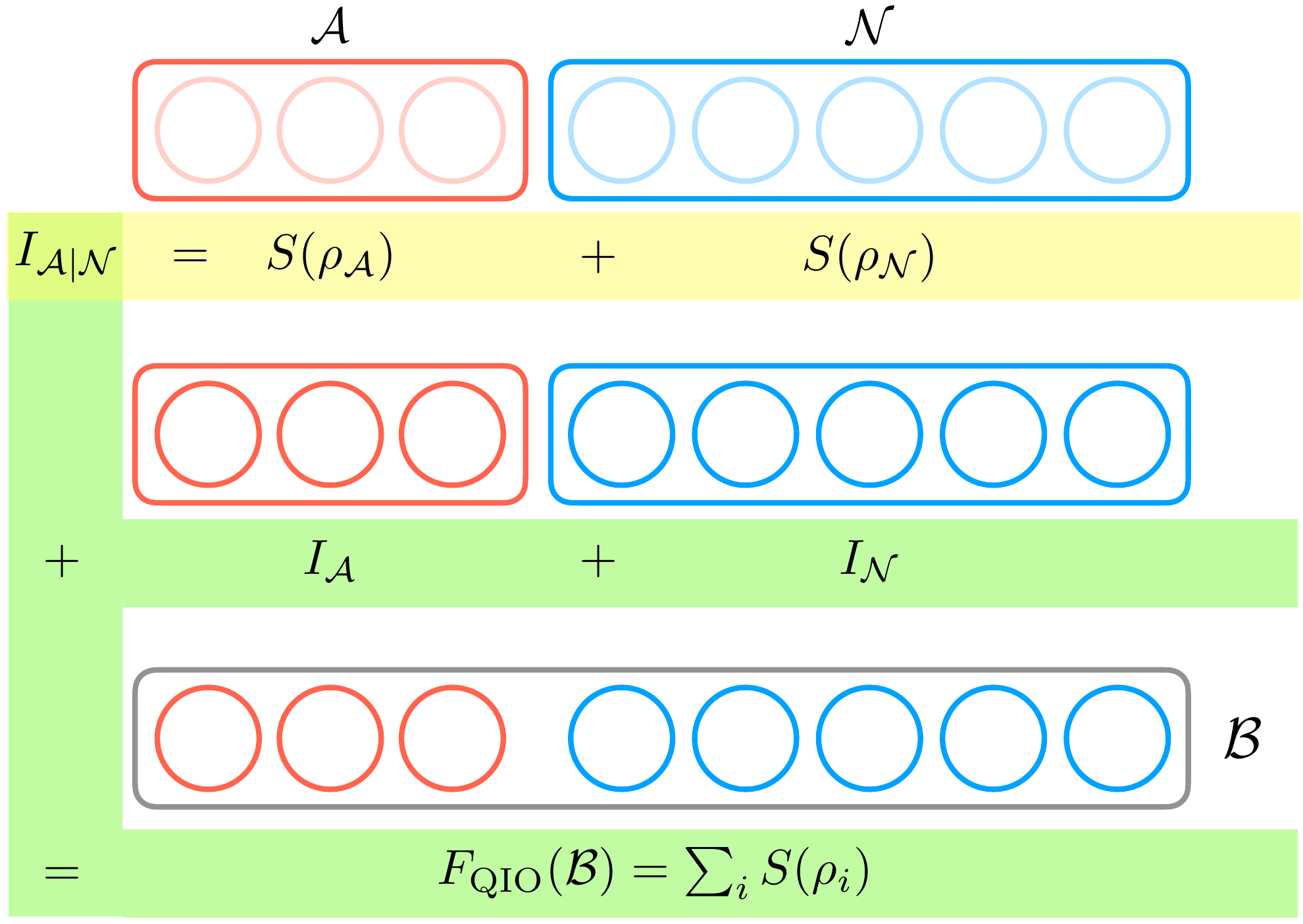}
\caption{Illustration of coarse-grained and finest-grained orbital correlation,
and the associating sum rules \eqref{eqn:fig2-1} (terms in yellow) and
\eqref{eqn:qio_cost_AvsN} (terms in green).}
\label{fig:sum_rule}
\end{figure}

To provide more evidence for the distinctive suitability of our cost function
\eqref{equ:qio_cost}, let us first consider instead of the finest split a
coarser partitioning $\Pi\equiv \mathcal{S}_1|\ldots| \mathcal{S}_R$ of the
orbital basis $\mathcal{B}$. As an extension of Eq.~\eqref{equ:I}, for any pure
quantum state $\rho$ the correlation between various subsystems $\mathcal{S}_J$
is quantified by
$
  I_{\Pi}(\rho) = \sum_{J =1}^R S(\rho_J).
$
Here $\rho_J$ is the orbital reduced state of subsystem $\mathcal{S}_J$
including all orbitals $i \in \mathcal{S}_J$. This coarse-grained correlation
vanishes indeed if  and only if the total state $\rho$ takes the form of a
product of reduced states $\rho_J$ of various subsystems $\mathcal{S}_J$. Yet,
$I_{\Pi}(\rho) $ is therefore \emph{not} capable of detecting any correlation
\emph{within} any of the subsystems $\mathcal{S}_J$, in striking contrast to our
cost function \eqref{equ:qio_cost} which refers to the finest partitioning of
$\mathcal{B}$.
%and $\rho_\mathcal{S}$ is the reduced state obtained by tracing out all orbital
%degrees of freedom outside of $\mathcal{S}$.
To put this into context, we consider the partition $\Pi =
\mathcal{A}|\mathcal{N}$ of the orbitals into active and non-active spaces
$\mathcal{A}$ and $\mathcal{N}$, respectively, as illustrated in the first row
of Figure \ref{fig:sum_rule}. The correlation between the two subspaces is
quantified as
\begin{equation}
  I_{\mathcal{A}|\mathcal{N}} = S(\rho_\mathcal{A}) + S(\rho_{\mathcal{N}}). \label{eqn:fig2-1}
\end{equation}
Additionally, the two subspaces $\mathcal{A}$ and $\mathcal{N}$ contain internal
correlations, when referring to the finest orbital partition within
$\mathcal{A}$ and $\mathcal{N}$ (second row of Figure \ref{fig:sum_rule}), which
are quantified as $I_{\mathcal{A}/\mathcal{N}} =
\sum_{i\in\mathcal{A}/\mathcal{N}} S(\rho_i) -
S(\rho_{\mathcal{A}/\mathcal{N}})$. Therefore, combining the external
correlation between $\mathcal{A}$ and $\mathcal{N}$, and the respective internal
correlation, we arrive at our cost function $F_\mathrm{QIO}$ which equals the
total orbital correlation referring to the finest orbital splitting (third row
in Figure \ref{fig:sum_rule})
\begin{equation}\label{eqn:qio_cost_AvsN}
  F_\mathrm{QIO}(\mathcal{B}) = \sum_i S(\rho_i) = I_{\mathcal{A}|\mathcal{N}} + I_\mathcal{A} + I_\mathcal{N}.
\end{equation}
Accordingly, as a key insight of our work, minimizing
$F_\mathrm{QIO}(\mathcal{B})$ means nothing else than reducing simultaneously
the correlation between the active space and the non-active space, as well as
within these two subspaces.

Also from a practical point of view, our cost function is particulary suitable.
To explain this, we first recall that in certain scenarios, for instance DMRG,
the entropy of blocks containing more than one orbital can be efficiently
exploited for orbital optimization due to the unique structure of the
ansatz~\cite{krumnow2016}. But, in general, going beyond the current partition
will require higher order RDMs, which is computationally expensive for ans\"atze
that do not possess the special structure of an MPS.

After having presented all these appealing features of our proposed cost
function \eqref{equ:qio_cost}, one may wonder which orbitals its minimization
will actually yield. Providing a comprehensive analytical answer is out of reach
due to the huge complexity of the electron correlation problem and the form of
\eqref{equ:qio_cost}. Yet, we recall a remarkable observation made in
Ref.~\cite{gigena2015a}. There, our cost function \eqref{equ:qio_cost} was
considered for spinless fermions, or equivalently for spinful fermions the total
correlation with respect to the finest splitting of the one-particle Hilbert
space into spin-orbitals. In that case, the minimization leads to the natural
spin-orbitals, i.e., the eigenstates of the full 1-RDM including
spin-information. In our case, however, the ideas of the derivation in
\cite{gigena2015a} do not apply anymore. Hence, the spatial orbitals minimizing
our cost function \eqref{equ:qio_cost}, termed quantum information-based
orbitals (QIOs), will not coincide with the natural orbitals but they could be
quite similar, at least for some systems. In that sense, our work may provide a
quite surprising alternative characterization of the natural orbitals: They
approximately but not exactly minimize the total orbital correlation in an
$N$-electron wavefunction. Because of this, we will also consider the natural
spatial orbitals in this work together with the QIOs obtained by minimizing
\eqref{equ:qio_cost}.

Based on the above considerations, we conclude that Eq~\eqref{equ:qio_cost} is
ideal for minimizing the representational complexity of the many-body
wavefunction from both a theoretical and practical point of view. In the
remainder of the Letter, we will use the term orbital to refer to spatial
orbitals only.

To turn the above theoretical insights into practical use, it is critical to
calculate the 1- and 2-RDM in a both efficient and accurate way in the whole
orbital space. Low-bond DMRG was used in previous
studies~\cite{stein2016,ding2023b} for this purpose. However, in practice this
approach becomes computationally expensive when the system size increases. In
this Letter, we propose to use the tailored CCSD ansatz
(TCCSD)~\cite{kinoshita2005}.
%where the active space is treated by a complete active space CI (CASCI) and the
%rest of the space is treated by restricted CCSD with inputs from the former.
This enables us to obtain the 1- and 2-RDM efficiently in the whole orbital
space, as well as incorporate static and dynamic correlations simultaneously. We
stress that TCCSD is used here as an example and the theoretical framework
offered by this work can be applied to other ans\"atze as well.

TCCSD~\cite{kinoshita2005} was introduced to overcome the shortcomings of CCSD,
namely its inadequacy in treating multireference systems, while retaining its
merits at capturing dynamic correlations. Here, we focus on TCCSD paired with
the complete active space CI (CASCI) solver and on the singlet ground state.
First, a FCI calculation in a predefined CAS space size of ($n^{\rm CAS}_{e},
n^{\rm CAS}_{o}$) is performed, where $n^{\rm CAS}_{e}$ and $n^{\rm CAS}_{o}$
are the number of electrons and spatial orbitals in the CAS space, respectively.
The CASCI solution $|\Psi_\mathrm{CAS}\rangle$ can be expanded in the
configurational basis up to a normalization prefactor as
\begin{equation}
\begin{split}
    |\Psi_\mathrm{CAS}\rangle &= c_0 |D_0\rangle + \sum_{i,a \in \mathcal{A}} c_i^a \hat{a}^\dagger_a \hat{a}^{\phantom{\dagger}}_i |D_0\rangle 
    \\
    &\quad+ \sum_{i,j,a,b \in \mathcal{A}} c_{ij}^{ab} \hat{a}^\dagger_a \hat{a}^\dagger_b \hat{a}^{\phantom{\dagger}}_j \hat{a}^{\phantom{\dagger}}_i |D_0\rangle + \cdots,
\end{split}
\end{equation}
where $|D_0\rangle$ is the reference determinant chosen by
occupying those orbitals which have the highest occupation numbers. In the initial step, it is chosen as the Hartree-Fock determinant.
In the TCCSD ansatz, the active space CI expansion coefficients $c_i^a$ and
$c_{ij}^{ab}$ are related to the $T_1^{\rm CAS}$ and $T_2^{\rm CAS}$ amplitudes
by the following expressions~\cite{kinoshita2005}
\begin{equation}
\begin{split}
    \hat{T}_1^{\mathrm{CAS}} &= \frac{1}{c_0} \sum_{i,a\in\mathcal{A}} c_i^a \hat{a}^\dagger_a \hat{a}^{\phantom{\dagger}}_i,
    \\
    \hat{T}_2^{\mathrm{CAS}} &= \frac{1}{c_0^2} \sum_{i,j,a,b\in \mathcal{A}} [c_0 c_{ij}^{ab} - (c_i^a c_j^b - c_i^b c_j^a)] \hat{a}^\dagger_a \hat{a}^\dagger_b \hat{a}^{\phantom{\dagger}}_j \hat{a}^{\phantom{\dagger}}_i.
\end{split}
\end{equation}
Then the TCCSD ansatz reads
\begin{equation}
    \ket{\Psi_{\rm TCCSD}} = \exp(\hat{T}_1^{\rm CAS} + \hat{T}_2^{\rm CAS} + \hat{T}_1^{\rm ext} + \hat{T}_2^{\rm ext}) \ket{D_0}, \label{eqn:tccsd}
\end{equation}
where $\hat{T}_1^{\mathrm{ext}} = \sum_{(i,a) \notin \mathcal{A}^2} T_i^a
\hat{a}^\dagger_a \hat{a}^{\phantom{\dagger}}_i$ and $\hat{T}_2^{\mathrm{ext}} =
\sum_{(i,j,a,b)\notin \mathcal{A}^4} T_{ij}^{ab} \hat{a}^\dagger_a
\hat{a}^\dagger_b \hat{a}^{\phantom{\dagger}}_j \hat{a}^{\phantom{\dagger}}_i$
contain the rest of the excitations. When solving the CCSD amplitudes equations,
$T_1^{\rm CAS}$ and $T_2^{\rm CAS}$ amplitudes are fixed and $T_1^{\rm ext}$ and
$T_2^{\rm ext}$ amplitudes ($T_i^a$ and $T_{ij}^{ab}$, respectively) are
optimized.

For the purpose of our orbital optimization scheme, we need to calculate the
single orbital entropy $S(\rho_i)$ from \eqref{eqn:tccsd}. The von Neumann
entropy $S$ is only well defined for orbital reduced states $\rho_i$'s that are
positive semi-definite, which is guaranteed if the 1- and 2-RDM, ${\gamma}$ and
${\Gamma}$, respectively, are computed symmetrically via the two- and four-point
correlation functions $\langle\Psi_\mathrm{TCCSD}|\hat{a}^{\dagger}_p
\hat{a}_q|\Psi_\mathrm{TCCSD}\rangle$ and
$\langle\Psi_\mathrm{TCCSD}|\hat{a}^{\dagger}_p\hat{a}^{\dagger}_q \hat{a}_s
\hat{a}_r|\Psi_\mathrm{TCCSD}\rangle$ with respect to the TCCSD wavefunction
\eqref{eqn:tccsd}. However, these expectations are indefeasibly expensive to
compute due to non-terminating commutators. To overcome this, normally the
$\Lambda$-CCSD theory is used to find the left eigenstate of the
similarity-transformed Hamiltonian. But the expectations now become asymmetric,
because the left and right eigenstates are not the same, and the positive
semi-definiteness of $\gamma$ and $\Gamma$ is lost in general. To allow for an
efficient computation while ensuring the positive semi-definiteness, we propose
to evaluate the 1- and 2-RDM as
\begin{equation}
  \begin{aligned}
  \gamma^{p}_{q} &= \braket{\Psi_{\rm CISD}|\hat{a}^{\dagger}_p \hat{a}_q|\Psi_{\rm CISD}}, \\
  \Gamma^{pq}_{rs} &= \braket{\Psi_{\rm CISD}|\hat{a}^{\dagger}_p \hat{a}^{\dagger}_q \hat{a}_s \hat{a}_r|\Psi_{\rm CISD}}, \\
  \end{aligned}
  \label{eqn:rdm_tccsd}
\end{equation}
where the expectations are with respect to $|\Psi_{\rm
CISD}\rangle=\frac{1}{C}[\ket{D_0}+\sum_{i,a} T_{i}^{a} \hat{a}^\dagger_a
\hat{a}^{\phantom{\dagger}}_i |D_0\rangle+\sum_{i,j,a,b}
(T_{ij}^{ab}+\frac{1}{2}T^a_iT^b_j) \hat{a}^\dagger_a \hat{a}^\dagger_b
\hat{a}^{\phantom{\dagger}}_j \hat{a}^{\phantom{\dagger}}_i |D_0\rangle]$,
normalized to 1 by $C$. The amplitudes used in $|\Psi_\mathrm{CISD}\rangle$ are
taken from the TCCSD ansatz, both active and non-active ones. The only caveat is
that \eqref{eqn:rdm_tccsd} neglects excitations in $|\Psi_\mathrm{TCCSD}\rangle$
beyond doubles. We stress that, however, the total energy and the amplitudes are
still solved for considering the full TCCSD formalism.  
%We notice in passing that it is also possible to obtain the RDMs in a similar
%fashion to the normal $\Lambda$-CCSD procedure~\cite{bartlett1989}. But in the
%TCCSD framework proper constraints on the active space amplitudes are needed to
%ensure the correct RDMs. We will report such a study in a future publication.
%We stress that this symmetric expression \eqref{eqn:rdm_tccsd} is crucial for
%maintaining the positive semi-definiteness of the orbital RDMs $\rho_i$'s in
%all orbital basis, so that the von Neumann entropy $S(\rho_i)$ is always
%well-defined.

\begin{figure}
    \centering
    \includegraphics[scale=1.2]{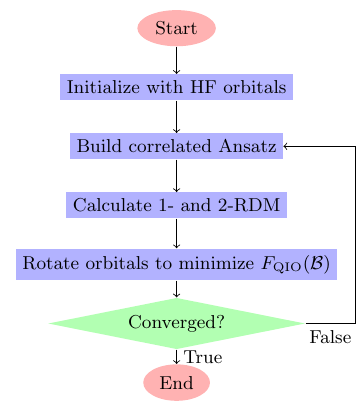}
    \caption{Flowchart of the QIO optimization algorithm. See the main text and the Appendix \ref{app:gradient} for more details.}
    \label{fig:qio_algo}
\end{figure}

\begin{figure}[ht!]
  \centering
  \includegraphics[scale=0.65]{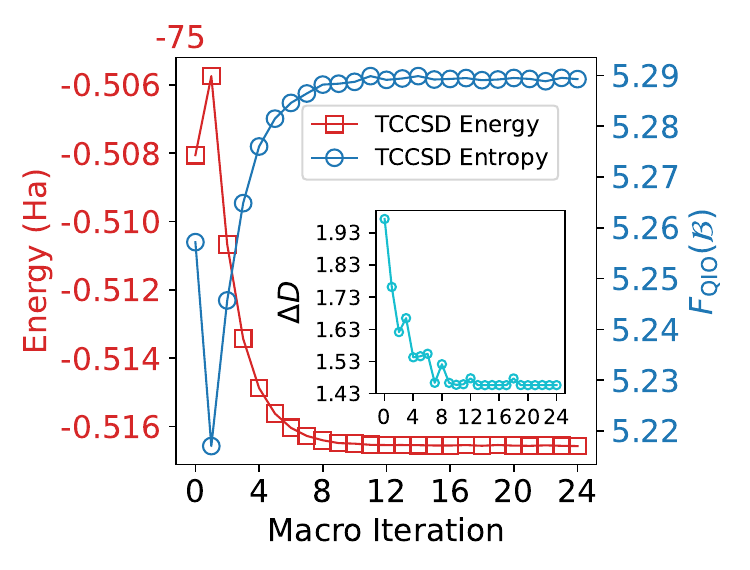}
  \caption{TCCSD-QIO optimization of C$_{\rm 2}$ at the bond length of 6.0 Bohr
  (3.175 ${\rm \AA}$) in the cc-pVDZ basis set. An active space size of (8,8) is
  used. The inset shows the distance between the TCCSD 2-RDM and that of a
  nearly exact DMRG calculation, defined as $\Delta D =
  \sqrt{\sum_{pqrs}|\Delta\Gamma^{pq}_{rs}|^2}$. The DMRG calculation is
  performed using the Block2 code~\cite{zhai2023}.}
   \label{fig:c2_demo}
  %The first two macro iterations are not shown, since the algorithm was at an
  %initial stage of looking for the reference determinant and TMP2 was used.}
\end{figure}

An iterative algorithm, consisting of macro and micro cycles as shown in
Fig.~\ref{fig:qio_algo}, is employed to find the optimal orbitals by minimizing
the cost function in Eq.~\eqref{equ:qio_cost}. The algorithm is initialized
normally with the HF orbitals at the beginning. The minimization of the cost
function is achieved by a quasi-Newton method, of which the details are
presented in Appendix \ref{app:gradient}. The convergence of the
overall algorithm is reached when the change in the total energy is smaller than
$10^{-6}$ Ha and the change in the cost function in the micro
cycle is smaller than $10^{-7}$.
In practice, we use a fixed number of micro cycles (20--30) in each macro cycle,
which within the initial macro cycles do not necessarily converge the cost function to the set
threshold and is designed so to avoid the whole optimization getting stuck in
local minima or saddle points. The overall convergence in both thresholds can be
achieved by increasing the total number of macro cycles.

For comparison, we
also carry out an iterative construction of the natural
orbitals, substituting the step where we minimize the cost function with the
step where we diagonalize the spin-traced 1-RDM to obtain NOs. In the case of
TCCSD, which relies on a dominant reference determinant, an additional step of
sorting the orbitals according to their occupation numbers decreasingly is
needed after the orbitals are rotated and before the correlated Ansatz is built
anew in the new orbital basis. The $N_{\rm act}$ orbitals around the Fermi
level, are then chosen as the active space orbitals after the orbitals are
sorted.

The Python code which achieves the QIO algorithm is available at~\cite{qio}. We
use PySCF~\cite{sun2020} for obtaining the initial HF orbitals, and for creating
the TCCSD algorithm. The DMRG calculations in this work are carried out by using
the Block2~\cite{zhai2023} package.

\begin{figure*}[ht!]
    \centering
  \includegraphics[width=\linewidth]{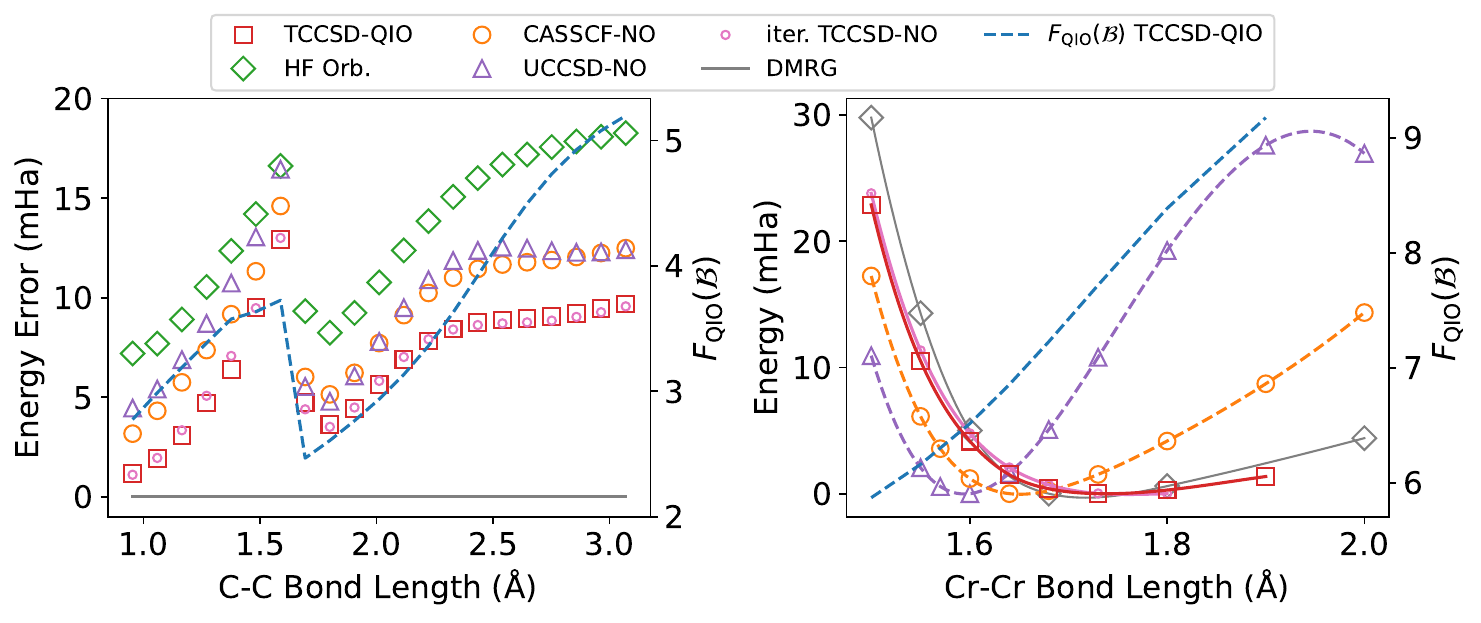}
  \caption{(Left) TCCSD energy errors relative to nearly exact DMRG
  results~\cite{wouters2014a} along the C$_{\rm 2}$ dissociation curve using
  various orbitals in the cc-pVDZ basis set. An active space size of (8,8) is
  used. (Right) Cr$_{\rm 2}$ potential energy curves, calculated by TCCSD and
  DMRG~\cite{larsson2022} in the cc-pVDZ-DK basis set. All energy curves are
  shifted by their respective lowest data points and are fitted using the cubic
  spline method. Entropies in the QIOs are shown in a blue dashed line.}
  \label{fig:c2_cr2_diss}
\end{figure*}

An example of the QIO algorithm in practice is shown in Fig.~\ref{fig:c2_demo}
for the orbital optimization of the C$_{\rm 2}$ at the bond length of 6.0 Bohr
(3.175 ${\rm \AA}$) in the cc-pVDZ basis set. In Appendix \ref{app:cr2} we present the analogous plot for the Cr$_{\rm 2}$ molecule. Only
macro iterations are shown, the micro iterations which minimize the cost
function $F_{\rm QIO}$ are not presented. An active space size of (8,8) is used.
Starting from the HF orbitals, the total orbital correlation $F_{\rm QIO}$
initially drops at the first macro iteration, then increases as the macro
iterations proceed. Correspondingly, the total TCCSD energy first increases and
then decreases. At first sight, it is counterintuitive that the total orbital
correlation increases as the macro iterations proceed, since the algorithm is
designed to minimize the total orbital correlation. Indeed, inside each macro
iteration, the total orbital correlation is minimized through orbital rotation,
while the quantum state is kept fixed. But in the next macro iteration, a new
correlated TCCSD wavefunction is built based on the set of updated orbitals by
solving again the amplitude equations, which allow for more correlation to be
captured in the ansatz, leading to a lower energy. We interpret this increase of
total orbital correlation on the level of macro iterations as a gain of
non-trivial information on the many-body wavefunction level, after trivial
information in the single-particle orbitals is compressed by the orbital
optimization. It is worth noting that, for a FCI solution, one would not need
the macro iterations. Because the FCI solution is exact in any basis and one can
find the best orbital basis by minimizing the total orbital correlation,
however, no more non-trivial information can be gained by doing so. Since TCCSD
is not variational, a lower energy cannot be taken as better for granted. In our
case, however, the lowering of the (non-variational) energy is indeed an
improvement, evident by the improved wavefunction quality quantified by the
distance between the TCCSD 2-RDM and that of a nearly exact DMRG calculation.
The distance between the two 2-RDMs is defined as
$
  \Delta D = \sqrt{\sum_{pqrs}|\Delta\Gamma^{pq}_{rs}|^2}%=\sqrt{\sum_i^M\sum_j^M\ket\Delta\Gamma_{ij}\ket^2}
$
where $\Delta\Gamma=\Gamma^{\rm TCCSD}-\Gamma^{\rm DMRG}$ and $q,p,r,s$ refer to
the indices of atomic orbitals. In the inset of Fig.~\ref{fig:c2_demo}, we show
that the distance indeed decreases as the macro iterations proceed, thus
justifying the final decreased TCCSD energy as an unmistakable improvement from
the HF-TCCSD energy.
%This is a strong evidence that the quality of the TCCSD ansatz is improved as
%the orbitals are optimized. To demonstrate the practicality of our QIO
%algorithm at handling a large number of orbitals, we refer to the SM for
%details on the optimization of orbitals of the Cr$_{\rm 2}$ molecule in the
%cc-pV5Z-DK basis set, consisting of in total 306 orbitals.

We now compare the TCCSD energy errors with various common choices of orbitals
along the dissociation curve of $\rm{C_2}$ in Fig.~\ref{fig:c2_cr2_diss} (Left).
The total orbital correlation in the optimized orbitals is also shown. All
calculations are performed in the cc-pVDZ basis set, the same as in the
reference DMRG calculations from Ref.~\cite{wouters2014a}. Overall, the errors
in the TCCSD energies using QIOs are smaller than those using HF orbitals,
CASSCF-NOs and spin-averaged unrestricted CCSD natural orbitals (UCCSD-NOs). As
suspected, the CASSCF-NOs, which provide the lowest active space energy, are not
the optimal orbitals for the total correlation. Remarkably, in a large region
along the dissociation curve, the improvement in the total energy from
CASSCF-NOs/UCCSD-NOs to QIOs is as substantial as that from HF orbitals to the
former. This highlights the importance of the simultaneous consideration of
static and dynamic correlations in orbital optimization. As expected, the
iterative TCCSD-NOs results follow the QIO's closely, with small deviations
around the equilibrium bond length. We observe that the total orbital
correlation increases as the bond length increases, until it suddenly drops at
the level crossing point, then increases again. We point out that the total
orbital correlation is a useful indicator of the complexity of the wavefunction.
For instance, it signals the level crossing point, but its value cannot be used
to predict the accuracy of the TCCSD energy directly, as shown in
Fig.~\ref{fig:c2_cr2_diss} (Left). At very stretched bond lengths, the total
orbital correlation is indeed larger than those at shorter bond lengths, in
agreement with the intuition that the wavefunction has more multireference
character at dissociation.

In Fig.~\ref{fig:c2_cr2_diss} (Right), we show the potential energy curves of
Cr$_{\rm 2}$ around the equilibrium point, calculated by TCCSD using different
orbitals along with the exact DMRG results~\cite{larsson2022} in the cc-pVDZ-DK
basis set. The relativistic effects are treated by using the scalar relativistic
exact two-component (X2C) Hamiltonian~\cite{kutzelnigg2005,peng2012} as
implemented in PySCF~\cite{sun2020}. A minimal active space of (12,12) is used
in TCCSD calculations. All curves are shifted by their respective lowest data
points. This system has attracted extensive studies over the
years~\cite{roos2003,kurashige2011,hongo2012,limanni2013,yamada2013,purwanto2015a,vancoillie2016,leszczyk2022a,larsson2022},
and serves as a benchmark for the performance of various methods at handling
strong correlations. Even around the equilibrium bond length, the ground state
exhibits strong multireference characters, and to capture the shallow potential
well requires also the inclusion of dynamic correlations. We note in passing
that a previous study employing TCCSD has shown rather unsatisfactory
results~\cite{leszczyk2022a}, i.e.~a too steep potential energy well, which
makes it a perfect test for QIOs.

First, we use UCCSD-NOs as done in DMRG~\cite{larsson2022}. With these orbitals,
TCCSD energies yield a significantly shorter equilibrium bond length of 1.595
${\rm \AA}$, compared to the exact DMRG result of 1.722 ${\rm \AA}$, with
increasingly larger errors as the bond length increases, resulting in a very
steep potential well. The TCCSD energies using CASSCF-NOs are better than using
UCCSD-NOs, with a predicted equilibrium bond length of 1.651 ${\rm \AA}$.
However, it yields a qualitatively wrong shape of the dissociation curve at
shorter and longer bond lengths, resulting still in a too steep potential well.
TCCSD in the QIOs, on the other hand, captures correctly the shallow potential
well around the equilibrium point and in general improves the curve both at
shorter and longer bond lengths. Especially, the equilibrium bond length
predicted by TCCSD using QIO is 1.737 ${\rm \AA}$, which agrees well with the
DMRG result. Most interestingly, we see a difference between the TCCSD-NOs and
TCCSD-QIOs results at slightly stretched bond lengths: the former does not
produce a bounded potential energy curve up until 1.8 ${\rm \AA}$, after which
the iterative TCCSD-NO algorithm becomes unstable and diverges. The difference
in the TCCSD energies between two stretched bond lengths is very small,
therefore this unbounded behavior is more evident when we examine the energies
values directly, listed in Appendix \ref{app:data}.

As noted in Ref.~\cite{larsson2022}, the ground state around 2.25~${\rm \AA}$ is
particularly difficult to capture even with very high bond dimensions in DMRG.
We also observe that starting from 2.0 ${\rm \AA}$, the TCCSD iterations in the
QIO optimizations do not converge to the threshold of $10^{-6}$ in energy and
the TCCSD energies oscillate between macro iterations. Therefore, we do not show
the TCCSD-QIO results at bond lengths larger or equal than  2.0 ${\rm \AA}$.
%We notice at 2.0 ${\rm \AA}$ the reference determinant retains a small
%coefficient around 0.28 in the CASCI calculation.
This likely reflects the limitations of the TCCSD ansatz with a minimal active
space to represent states with very strong multireference character rather than
potential shortcomings of the QIO algorithm itself. The entropy of the QIOs is
shown as a blue dashed line in Fig.~\ref{fig:c2_cr2_diss} (Right). Its smooth
increase with the bond length hints at the increasing multireference character
of the wavefunction, and serves as an indicator if unphysical results are
obtained in the TCCSD calculations when going to larger bond lengths. As pointed
out in previous studies on Cr$_{\rm
2}$~\cite{limanni2013,purwanto2015a,larsson2022}, to capture qualitatively
correct the whole potential energy curve, a larger active space is needed.

We introduced a quantum information-based orbital (QIO) optimization scheme,
emerging from a fresh perspective on one of the crucial steps --- orbital
optimization --- in solving the strongly correlated many-body problem. The QIOs
are characterized as those orbitals that minimize the total orbital correlation.
Accordingly, they reveal the intrinsic complexity of the many-body wavefunction
by compressing all trivial information into single-particle orbitals. Due to its
distinctive nature, our scheme addresses the challenging task of concurrently
treating both static and dynamic correlations in strongly correlated systems by
optimizing orbitals considering both types of correlations. By employing our
orbital optimization scheme within the TCCSD framework, we obtained superior
results compared to other commonly used orbitals. The iterative orbital
optimization also provides, to some extent, self-consistent feedback between the
active and external space treated separately by CASCI and CCSD. This addresses
partially the long-standing issue of the lack of self-consistency in the TCCSD
framework. Through the lens of total orbital correlation, we also explained and
demonstrated the close relationship between the QIOs and iteratively constructed
NOs. The QIOs use information from both the 1- and 2-RDM, where the latter can
be decomposed as $\Gamma^{pq}_{rs}=\lambda^{pq}_{rs} +
\gamma^p_q\gamma^r_s-\gamma^p_s\gamma^r_q$ and $\lambda^{pq}_{rs}$ is the
two-body cumulant~\cite{kutzelnigg1999,mazziotti1998a}. The $\lambda^{pq}_{rs}$
quantifies the non-trivial two-body correlation in a system. So we suspect that
in more challenging cases where the two-body cumulant is non-negligible, these
two types of orbitals will differ more significantly, hinted by the Cr$_{\rm 2}$
case. The computational cost of the QIO algorithm depends on the ansatz used. In
the case of TCCSD, the computational cost can be broken down into three parts:
i) the CASCI calculation in the active space, formally scales exponentially with
the active space size; ii) the TCCSD calculation, which scales polynomially
$N^6$, where $N$ is the number of electrons;; and iii) the orbital optimization,
which is dominated by the rotation of the 2-RDM and scales $M^5$, where $M$ is
the number of orbitals. The current bottleneck in TCCSD is the limited active
space size that can be treated by CASCI, which can straightforwardly be overcome
by using DMRG~\cite{veis2016} or FCIQMC~\cite{vitale2020} as the active space
solver. In the future, we hope to extend this algorithm to other systematically
improvable methods, such as auxiliary-field quantum Monte Carlo
(AFQMC)~\cite{purwanto2015a,lee2022,shi2021}, as well as to spin states beyond
singlets. Given the demonstrated advantages of QIOs and the comparable
computational cost of QIO to CASSCF, we believe that the QIO algorithm may lead
to a change of paradigm in how we approach strongly correlated many-electron
systems.

\textit{Acknowledgement.} We thank Huanchen Zhai and Seunghoon Lee for useful
discussions on DMRG and TCCSD. We are also grateful to Ali Alavi, Daniel Kats,
Stefan Knecht and Giovanni Li Manni for comments. We acknowledge financial
support from the German Research Foundation (Grant SCHI 1476/1-1), the Munich
Center for Quantum Science and Technology, and the Munich Quantum Valley, which
is supported by the Bavarian state government with funds from the Hightech
Agenda Bayern Plus.

\appendix

\section{Analytic Gradients, Hessians and Quasi-Newton Algorithm}\label{app:gradient}
The objective here is to minimize the sum of orbital entropies of all orbitals, amounting to the following cost function
\begin{equation}
  \fqi(\mathcal{B}) = \sum_{i=1}^D S(\rho_i)
\end{equation}
Let $\mathbf{B}$ be the matrix of molecular orbital coefficients where each column corresponds to a molecular orbital. Then the transformation from the initial orbital basis $\mathbf{B}_0$ to the minimizing orbital basis $\mathbf{B}_\mathrm{opt}$ can be achieved by a unitary transformation $\mathbf{U}_\mathrm{opt} = \exp(\mathbf{X}_\mathrm{opt})$ as
\begin{equation}
  \mathbf{B}_\mathrm{opt} = \mathbf{B}_0 \mathbf{U}^\dagger_\mathrm{opt}.
\end{equation}

With the initial orbital basis fixed, e.g. the HF basis, we can denote all orbital bases by the corresponding real-value unitary $\mathbf{U}$ or its generator, namely a skew-symmetric matrix $\mathbf{X}$. The set of skew-symmetric matrices can be parameterized as
\begin{eqnarray}
  \mathbf{X} = \sum_{i<j} x_{ij} \mathbf{A}^{(ij)}
\end{eqnarray}
where $\mathbf{A}^{(ij)}$ is an antisymmetric matrix with elements $A^{(ij)}_{ij}=-A^{(ij)}_{ji}=1$ and otherwise 0. The cost function can be reparameterized as
\begin{eqnarray}
  \fqi(\mathcal{B}) \equiv \fqi(\mathbf{B}) = \fqi(\mathbf{B}_0(e^{\mathbf{X}})^\dagger) \equiv \fqi(\mathbf{X}).
\end{eqnarray}
At a local extremum, $\fqi(\mathbf{X}_\mathrm{opt})$ satisfies the following extremal conditions
\begin{equation}
  \begin{split}
  &\quad\frac{\partial \fqi}{\partial x_{ij}}(\mathbf{X}_\mathrm{opt})
  \\
  &=\lim_{\theta \rightarrow 0} \frac{\fqi(\mathbf{X}_\mathrm{opt}\!+\!\theta \mathbf{A}^{(ij)})\!-\!\fqi(\mathbf{X}_\mathrm{opt})}{\theta} = 0,
  \end{split}
\end{equation}
for all tuples $(i,j)$ such that $i<j$. In practice, the translation invariance of the underlying $D(D-1)/2$ dimensional real space can be exploited to simplify the exponential. To be more specific, one can set the current matrix $\mathbf{X}$ to be the origin at every gradient step, which is the same as constantly updating the reference orbital basis $\mathbf{B}_0$. In that case, the derivative of the cost function simplifies to
\begin{eqnarray}
  \frac{\partial \fqi}{\partial x_{ij}}(\mathbf{0}) = \lim_{\theta \rightarrow 0} \frac{\fqi(\theta \mathbf{A}^{(ij)})\!-\!\fqi(\mathbf{0})}{\theta}.
\end{eqnarray}
The infinitesimal unitary transformation $\mathbf{J}^{(ij)}(\theta)=e^{\theta\mathbf{A}^{(ij)}}$ associating to this derivative is simply a Jacobi rotation between orbital $i$ and $j$:
\begin{equation}
\begin{split}
  (\mathbf{J}^{(ij)}(\theta))_{ii} &= \phantom{-}\cos(\theta),
  \\
  (\mathbf{J}^{(ij)}(\theta))_{ij} &= \phantom{-}\sin(\theta),
  \\
  (\mathbf{J}^{(ij)}(\theta))_{ji} &= -\sin(\theta),
  \\
  (\mathbf{J}^{(ij)}(\theta))_{jj} &= \phantom{-}\cos(\theta),
\end{split}
\end{equation}
and $(\mathbf{J}^{(ij)}(\theta))_{kl}=\delta_{kl}$ for $k,l\neq i,j$. The derivatives of $\fqi$ can be further broken down using the chain rule
\begin{equation}
  \begin{split}
    \frac{\partial \fqi}{\partial x_{ij}}(\mathbf{0}) = -\sum_{i=1}^D\sum_{k=0}^{3} \log(\lambda_k^{(i)}(\mathbf{0})) \frac{\partial \lambda_k^{(i)}}{\partial x_{ij}}(\mathbf{0}),
  \end{split}
  \label{eqn:first_derivative}
\end{equation}
where
\begin{equation}
  \begin{split}
    \lambda_0^{(i)}(\mathbf{X}) &= 1 - \gamma(\mathbf{X})^{i}_{i} - \gamma(\mathbf{X})^{\bar{i}}_{\bar{i}} + \Gamma(\mathbf{X})^{i\bar{i}}_{i\bar{i}},
    \\
    \lambda_1^{(i)}(\mathbf{X}) &= \gamma(\mathbf{X})^{i}_{i}  - \Gamma(\mathbf{X})^{i\bar{i}}_{i\bar{i}},
    \\
    \lambda_2^{(i)}(\mathbf{X}) &= \gamma(\mathbf{X})^{\bar{i}}_{\bar{i}} - \Gamma(\mathbf{X})^{i\bar{i}}_{i\bar{i}},
    \\
    \lambda_3^{(i)}(\mathbf{X}) &= \Gamma(\mathbf{X})^{i\bar{i}}_{i\bar{i}}.
  \end{split}
\end{equation}
Here, $\gamma(\mathbf{X})$ and $\Gamma(\mathbf{X})$ are the 1- and 2-RDMs in the rotated basis given by
\begin{equation}
  \begin{split}
    \gamma(\mathbf{X})^{i}_{i} &= \sum_{ab} (e^{\mathbf{X}})_{ia}(e^{\mathbf{X}})_{ib} \gamma(\mathbf{0})^{a}_{b},
    \\
    \Gamma(\mathbf{X})^{i\bar{i}}_{i\bar{i}} &= \sum_{abcd} (e^{\mathbf{X}})_{ia}(e^{\mathbf{X}})_{ib}(e^{\mathbf{X}})_{ic}(e^{\mathbf{X}})_{id} \Gamma(\mathbf{0})^{a\bar{b}}_{c\bar{d}},
  \end{split}
\end{equation}
Eventually, their derivatives at $\mathbf{0}$ can be computed from the partial derivatives of the unitary
\begin{equation}
    \left.\frac{\partial \exp(\mathbf{X})}{\partial x_{ij}}\right|_{\mathbf{X}=\mathbf{0}} = \lim_{\theta \rightarrow 0} \frac{\mathbf{J}^{(ij)}(\theta)-\mathbbm{1}}{\theta} = (\mathbf{J}^{(ij)})'({0})
\end{equation}
where
\begin{equation}
  \begin{split}
    (\mathbf{J}^{(ij)})'({0})_{ii} &= \phantom{-}0,
    \\
    (\mathbf{J}^{(ij)})'({0})_{ij} &= \phantom{-}1,
    \\
    (\mathbf{J}^{(ij)})'({0})_{ji} &= -1,
    \\
    (\mathbf{J}^{(ij)})'({0})_{jj} &= \phantom{-}0,
  \end{split}
\end{equation}
with $(\mathbf{J}^{(ij)}(\theta))_{kl}=0$ for $k,l\neq i,j$. With this we can write the derivatives of the relevant entries of the RDMs as
\begin{equation}
  \begin{split}
    \frac{\partial \gamma(\mathbf{0})^i_{i}}{\partial x_{ij}} &= \gamma(\mathbf{0})_{i}^j + \gamma(\mathbf{0})_{j}^i,
    \\
    \frac{\partial \gamma(\mathbf{0})^j_{j}}{\partial x_{ij}} &= -\gamma(\mathbf{0})_i^j - \gamma(\mathbf{0})_j^i,
    \\
    \frac{\partial \Gamma(\mathbf{0})^{i\bar{i}}_{i\bar{i}}}{\partial x_{ij}} &= \Gamma(\mathbf{0})^{j\bar{i}}_{i\bar{i}} + \Gamma(\mathbf{0})^{i\bar{j}}_{i\bar{i}} + \Gamma(\mathbf{0})^{i\bar{i}}_{j\bar{i}} + \Gamma(\mathbf{0})^{i\bar{i}}_{i\bar{j}}.
    \\
    \frac{\partial \Gamma(\mathbf{0})^{j\bar{j}}_{j\bar{j}}}{\partial x_{ij}} &= -\Gamma(\mathbf{0})^{i\bar{j}}_{j\bar{j}} - \Gamma(\mathbf{0})^{j\bar{i}}_{j\bar{j}} - \Gamma(\mathbf{0})^{j\bar{j}}_{i\bar{j}} - \Gamma(\mathbf{0})^{j\bar{j}}_{i\bar{j}}.
  \end{split}
\end{equation}
For the quasi-Newton algorithm we also need the second derivative of the cost function, which we shall approximate with its diagonal elements
\begin{equation}
  \begin{split}
  \frac{\partial^2 F_{\mathbf{QI}}(\mathrm{0})}{\partial x_{ij}^2} = &-\sum_{i=1}^{D}\sum_{k=0}^{3}\left[  \frac{1}{\lambda_k^{(i)}(\mathbf{0})} \left(\frac{\partial\lambda_k^{(i)}(\mathbf{0})}{\partial x_{ij}}\right)^2 \right.
  \\
  &+ \left.\log(\lambda^{(i)}_k(\mathbf{0}))\frac{\partial^2 \lambda_k^{(i)}(\mathbf{0})}{\partial x_{ij}^2}\right],
  \end{split}
  \label{eqn:second_derivative}
\end{equation}
which involves the second derivatives of the RDMs. Again, from the second derivative of the orbital rotation matrix
\begin{equation}
  \begin{split}
    (\mathbf{J}^{(ij)})''({0})_{ii} &= -1,
    \\
    (\mathbf{J}^{(ij)})''({0})_{ij} &= \phantom{-}0,
    \\
    (\mathbf{J}^{(ij)})''({0})_{ji} &= \phantom{-}0,
    \\
    (\mathbf{J}^{(ij)})''({0})_{jj} &= -1,
  \end{split}
\end{equation}
with $(\mathbf{J}^{(ij)}(\theta))_{kl}=0$ for $k,l\neq i,j$, we can derive the second derivatives of the relevant entries of the RDMs
\begin{equation}
  \begin{split}
    \frac{\partial^2 \gamma(\mathbf{0})^i_{i}}{\partial x_{ij}^2} &= -2\gamma(\mathbf{0})_{i}^i + 2\gamma(\mathbf{0})_{j}^j,
    \\
    \frac{\partial^2 \gamma(\mathbf{0})^j_{j}}{\partial x_{ij}^2} &= 2\gamma(\mathbf{0})_{i}^i - 2\gamma(\mathbf{0})_{j}^j,
    \\
    \frac{\partial^2 \Gamma(\mathbf{0})^{i\bar{i}}_{i\bar{i}}}{\partial x_{ij}^2} &= -4\Gamma(\mathbf{0})^{i\bar{i}}_{i\bar{i}} + 2 \sum_{(a,b,c,d)\in \mathcal{P}(i,i,j,j)} \Gamma(\mathbf{0})^{b\bar{d}}_{a\bar{c}},
    \\
    \frac{\partial^2 \Gamma(\mathbf{0})^{j\bar{j}}_{j\bar{j}}}{\partial x_{ij}^2} &= -4\Gamma(\mathbf{0})^{j\bar{j}}_{j\bar{j}} + 2 \sum_{(a,b,c,d)\in \mathcal{P}(i,i,j,j)} \Gamma(\mathbf{0})^{b\bar{d}}_{a\bar{c}},
  \end{split}
\end{equation}
where $\mathcal{P}(i,i,j,j)$ collects all permutations of the tuple $(i,i,j,j)$.

At each micro-iteration, the ${\bf X}$ matrix is updated by the following equation
\begin{equation}
  {\bf X} \leftarrow {\bf X} - \alpha [{\bf H}-({\rm min}({\bf H})+\delta)]^{-1} {\bf G},
\end{equation}
where $\alpha$ (typically $0.1-0.3$) is the step size, ${\bf H}$ is the diagonal Hessian matrix of the cost function~\eqref{eqn:second_derivative}, $\delta$ is a small positive level-shift
parameter ($10^{-4}-10^{-1}$) and ${\bf G}$ is the gradient of the cost function~\eqref{eqn:first_derivative}.

\section{Two-Electron Singlet State in Two Orbitals} \label{app:2el}

Let $|\Psi\rangle$ be a two-electron singlet state in two orbitals (which form a basis $\mathcal{B}$ of the orbital one-particle Hilbert space) associated with annihilation (creation) operators $f^{(\dagger)}_{1/2\sigma}$. Then the following form is general
\begin{equation}
  |\Psi(\mathbf{p})\rangle = p_0 |\Psi_0\rangle + p_1 |\Psi_1\rangle + p_2 |\Psi_2\rangle,
\end{equation}
where
\begin{equation}
  \begin{split}
    &|\Psi_0\rangle = f^\dagger_{1\uparrow}f^\dagger_{1\downarrow}|0\rangle,
    \\
    &|\Psi_1\rangle = \frac{1}{\sqrt{2}}(f^\dagger_{1\uparrow}f^\dagger_{2\downarrow}-f^\dagger_{1\downarrow}f^\dagger_{2\uparrow})|0\rangle,
    \\
    &|\Psi_2\rangle = f^\dagger_{2\uparrow}f^\dagger_{2\downarrow}|0\rangle.
  \end{split}
\end{equation}
Although the form of $|\Psi(\mathbf{p})\rangle$ is the most general, it is not the most concise in terms of its CI expansion, whose complexity can be measured by the Shannon entropy of the absolute squares of the CI coefficients
\begin{equation}
  \begin{split}
  &\quad H\left(\left\{p_0^2,\frac{p_1^2}{2},\frac{p_1^2}{2},p_2^2\right\}\right)
  \\
  &= -p_0^2\log(p_0^2) - p_1^2 \log\left(\frac{p_1^2}{2}\right) - p_2^2 \log(p_2).
  \end{split}
\end{equation}
In this very special case, the CI entropy precisely coincide with the entanglement between the two spatial orbitals, given by the von Neumann entropy
\begin{equation}
  S(\rho) = -\mathrm{Tr}[\rho\log(\rho)]
\end{equation}
of one of the orbital reduced density matrix
\begin{equation}
  \rho_{1/2}(\mathbf{p}) = \mathrm{Tr}_{2/1}[|\Psi(\mathbf{p})\rangle\langle\Psi(\mathbf{p})] = \begin{pmatrix}
    p_0^2 & 0 & 0 & 0
    \\
    0 & p_1^2/2 & 0 & 0
    \\
    0 & 0 & p_1^2/2 & 0
    \\
    0 & 0 & 0 & p^2_2
  \end{pmatrix}.
\end{equation}
Clearly, $S(\rho_{1/2}(\mathbf{p}))$ depends on the vector $\mathbf{p}$, which for a fixed state $|\Psi\rangle$ again depends on the orbital basis $\mathbf{B}$. We now determine the orbital basis in which the single orbital entropy as well as the multireference character of the state is minimized/maximized.

We consider all possible real orbital basis, which can be realized by a 2-by-2 orthogonal transformation from the current basis
\begin{equation}
  \begin{split}
  f^{(\dagger)}_{1\uparrow/\downarrow} &\longmapsto \cos(\theta) f^{(\dagger)}_{1\uparrow/\downarrow} - \sin(\theta)f^{(\dagger)}_{2\uparrow/\downarrow},
  \\
  f^{(\dagger)}_{2\uparrow/\downarrow} &\longmapsto \sin(\theta) f^{(\dagger)}_{1\uparrow/\downarrow} + \cos(\theta)f^{(\dagger)}_{2\uparrow/\downarrow}.
  \end{split}
\end{equation}
In the new orbital basis the state $|\Psi\rangle$ becomes
\begin{equation}
  \begin{split}
    |\Psi(\mathbf{p},\theta)\rangle = q_0(\theta) |\Psi_0\rangle + q_1(\theta) |\Psi_1\rangle + q_2(\theta) |\Psi_2\rangle,
  \end{split}
\end{equation}
where the new coefficients are given as
\begin{equation}
  \begin{split}
    q_0(\theta) &= p_0\cos^2(\theta)+\sqrt{2}p_1\cos(\theta)\sin(\theta) + p_2 \sin^2(\theta),
    \\
    q_1(\theta) &= \frac{1}{\sqrt{2}}\sin(2\theta)(p_2-p_0) + \frac{1}{\sqrt{2}}\cos(2\theta)p_1,
    \\
    q_2(\theta) &= p_0\sin^2(\theta) - \sqrt{2}p_1\cos(\theta)\sin(\theta) + p_2 \cos^2(\theta).
  \end{split}
\end{equation}

First of all, we notice that $q_1(\theta)$ can always be set to 0 for some choice of $\theta$. Therefore we can without loss of generality always assume that there is a basis such that the amplitude of $|\Psi_1\rangle$ is 0 and $p_0,p_2>0$. That is, a general singlet state becomes
\begin{equation}
  |\Psi(\mathbf{p})\rangle = p_0 |\Psi_0\rangle + p_2 |\Psi_2\rangle. \label{eqn:singlet}
\end{equation}
We will use this orbital basis as the reference basis, and the transformed amplitudes simplify to
\begin{equation}
  \begin{split}
    q_0(\theta) &= p_0\cos^2(\theta)+p_2 \sin^2(\theta),
    \\
    q_1(\theta) &= \sqrt{2}\cos(\theta)\sin(\theta)(p_2-p_0),
    \\
    q_2(\theta) &= p_0\sin^2(\theta)+ p_2 \cos^2(\theta).
  \end{split}
\end{equation}
Specially, $p_0=p_2=1/\sqrt{2}$ is a stationary point.

Second, the orbital entropy is minimal in the reference orbital basis, where the eigenvalues of the orbital RDMs are simply $\{p_0^2, p_2^2,0,0\}$, and the minimal entropy is given by
\begin{equation}
  S(\rho_{1/2}(\mathbf{p})) = -p_0^2\log(p_0^2) - p_2^2 \log(p_2^2).
\end{equation}
In a transformed orbital basis, the spectrum of the orbital RDMs is
\begin{equation}
\begin{split}
  \mathrm{Spec}(\rho_{1/2}(\mathbf{p},\theta)) = \left\{q_0(\theta)^2, \frac{q_1(\theta)^2}{2},\frac{q_1(\theta)^2}{2},q_2(\theta)^2\right\}.
\end{split}
\end{equation}
The largest eigenvalue of the RDMs is given by $\max{q_0(\theta)^2,q_2(\theta)^2}$, since
\begin{equation}
  \begin{split}
    \max\{q_0(\theta),q_2(\theta)\} &\geq \frac{q_0(\theta)+q_2(\theta)}{2}
    \\
    &= \frac{p_0(\theta)+p_2(\theta)}{2}
    \\
    &\geq \left|\frac{q_1(\theta)}{\sqrt{2}}\right|.
  \end{split}
\end{equation}
Additionally, it is easy to see that $p_0^2 \geq \max{q_0(\theta)^2,q_2(\theta)^2}$. We can therefore easily conclude that the spectrum of the RDMs at $\theta=0$ majorizes all other possible spectra, and that when $\theta=0$ the orbital entropy is at its lowest.

\begin{figure}[t]
  \centering
  \includegraphics[scale=0.32]{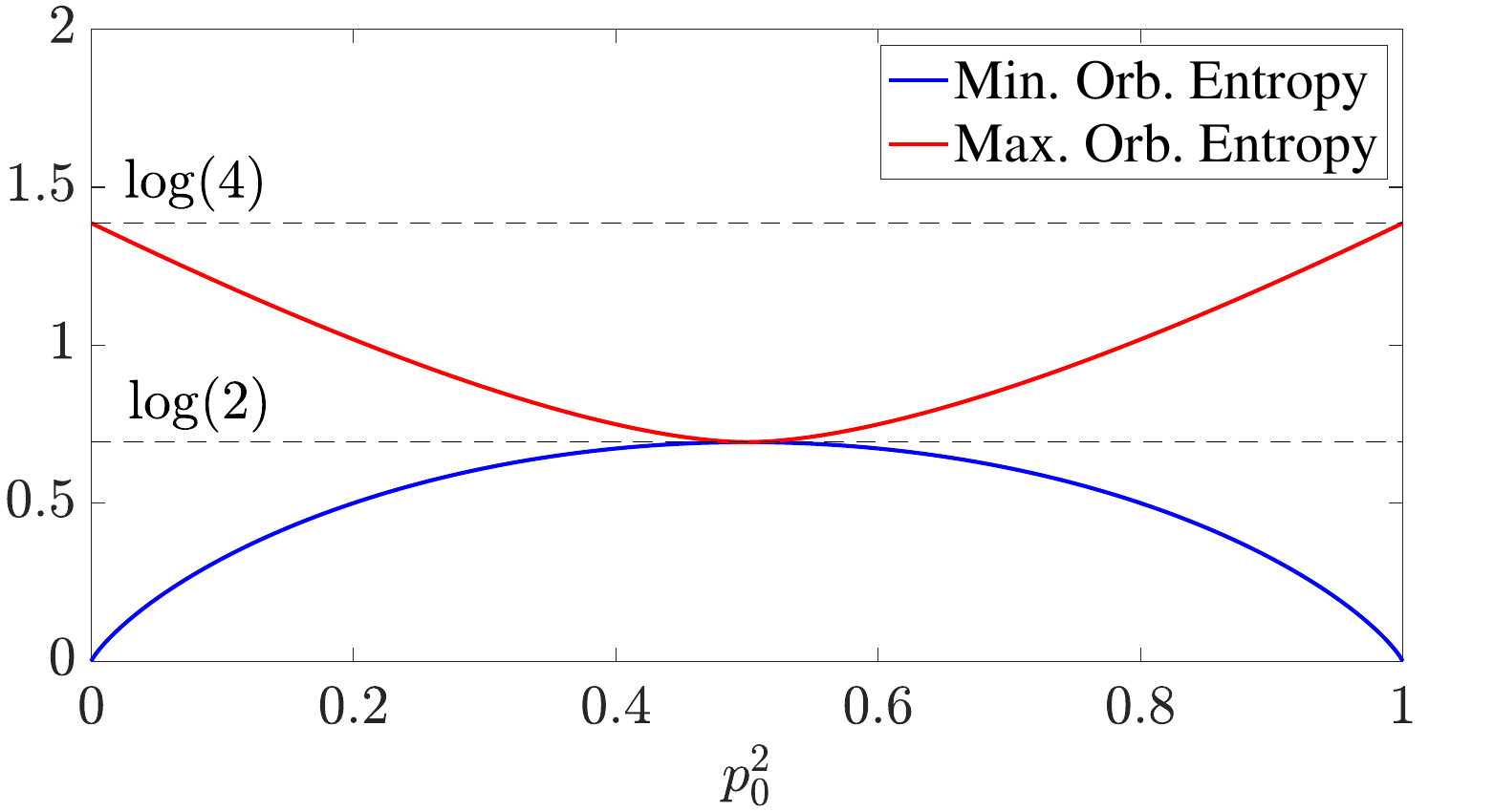}
  \caption{Minimal/maximal orbital entropy $S(\rho_{1/2})$ over all real orbital basis, against the amplitude $p_0^2$ in \eqref{eqn:singlet}.}
  \label{fig:2el}
\end{figure}

Third, maximal orbital entropy is achieved by a $\frac{\pi}{4}$-rotation from the reference orbitals. When $\theta=\frac{\pi}{4}$, the spectrum of the RDMs is given by
\begin{equation}
  \begin{split}
  \mathrm{Spec}\left(\rho_{1/2}\left(\mathbf{p},\frac{\pi}{4}\right)\right) &= \left\{ \left(\frac{p_0+p_2}{2}\right)^2,\left(\frac{p_0-p_2}{2}\right)^2,\right.
  \\
  &\left.\quad \left(\frac{p_0-p_2}{2}\right)^2,\left(\frac{p_0+p_2}{2}\right)^2 \right\}.
  \end{split}
\end{equation}
Notice that
\begin{equation}
  \begin{split}
    \mathrm{Spec}(\rho_{1/2}(\mathbf{p},\theta)) &\succ \left\{\frac{q_0(\theta)^2+q_2(\theta)}{2}, \frac{q_1(\theta)^2}{2},\right.
    \\
    &\quad \left.\frac{q_1(\theta)^2}{2},\frac{q_0(\theta)^2+q_2(\theta)}{2}\right\}
    \\
    &\succ \mathrm{Spec}\left(\rho_{1/2}\left(\mathbf{p},\frac{\pi}{4}\right)\right).
  \end{split}
\end{equation}
Therefore the orbital entropy is maximal when $\theta=\frac{\pi}{4}$, which equals to
\begin{equation}
  \begin{split}
  S\left(\rho_{1/2}\left(\mathbf{p},\frac{\pi}{4}\right)\right) &= - 2 \left(\frac{p_0+p_2}{2}\right)^2 \log\left[\left(\frac{p_0+p_2}{2}\right)^2\right]
  \\
  & \quad - 2 \left(\frac{p_0-p_2}{2}\right)^2 \log\left[\left(\frac{p_0-p_2}{2}\right)^2\right].
  \end{split}
\end{equation}

To summarize, through orbital rotation, one can minimize/maximize the orbital entropy of the two-electron state, which in this case coincide with a measure of multireference character, namely the CI entropy. The orbital entropy is minimized when $|\Psi\rangle$ is expressed in a zero seniority form, and is maximized when a $\frac{\pi}{4}$-rotation is applied to the minimizing orbital basis. These findings are encapsulated in Figure \ref{fig:2el}, where we presented the minimal/maximal single orbital entropy $S(\rho_{1/2})$ over all real orbital basis, against the amplitude $p_0^2$. When $p_0^2 = 0,1$, the state is a single Slater determinant, and correspondingly the minimal orbital entropy vanishes. As $p_0^2$ approaches $1/2$, $|\Psi\rangle$ becomes more multireference, and accordingly the minimal orbital entropy increases. When $p_0^2=1/2$, the two orbitals are equally correlated/entangled in every orbital basis. In other words, the complexity of the state cannot be transformed away by orbital rotation. In contrast, the maximal orbital entropy behaves in the opposite manner. It is maximal when $p_0^2 = 0,1$, and minimal when $p_0^2=1/2$. It signals the triviality of the orbital correlation of a single reference state in a poorly chosen orbital basis.

\begin{figure}[t]
  \centering
  \includegraphics[width=\columnwidth]{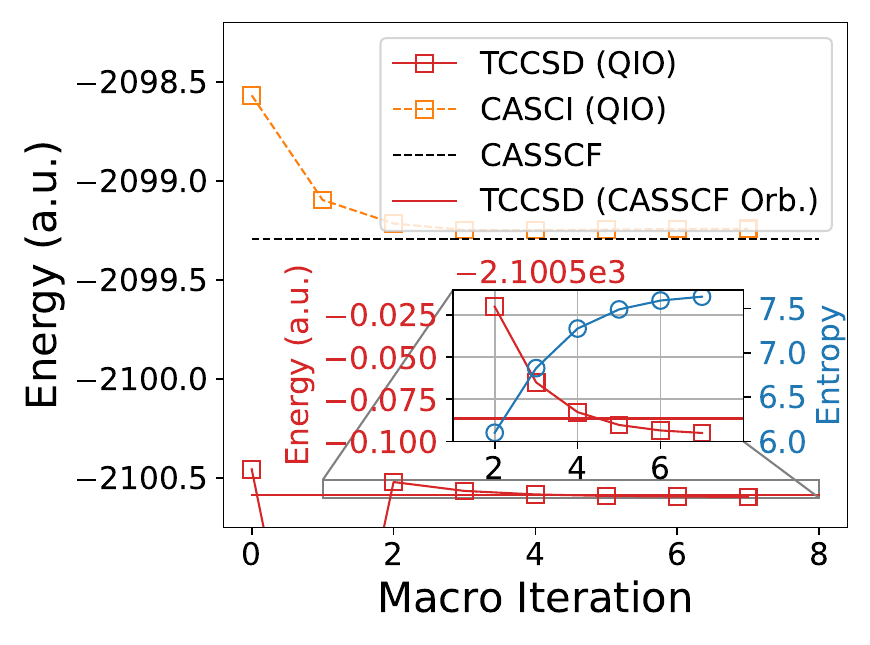}
  \caption{TCCSD-QIO optimization of the Cr$_{\rm 2}$  molecule at the bond length of 1.679 ${\rm \AA}$ in the cc-pV5Z-DK basis set, consisting of 306 orbitals.
  A minimal active space of (12,12) is used. The inset shows the details of the TCCSD energies and the total orbital correlation as
  the orbitals are optimized.}
  \label{fig:cr2_orb_opt}
\end{figure}

\section{TCCSD-QIO optimization of Cr$_{\rm 2}$}\label{app:cr2}

In the original QICAS paper~\cite{ding2023b},
a truncated set of orbitals were used due to the huge computational cost of low-bond DMRG calculation.
We also point out that the CASSCF algorithm in this case also gets very expensive.
%We will briefly comment on the computational cost of the QIO algorithm in the Conclusion section.
As the optimization proceeds,
the active space CASCI energies decrease and eventually converge slightly above the CASSCF energy,
while the TCCSD energies decrease below the one using CASSCF canonical orbitals.
The abrupt decrease in the TCCSD energy at iteration 1 results from the orbital rotation traversing a region where the reference determinant isn't dominant,
and the tailored MP2 (TMP2) ansatz is employed to navigate out of this region.
%In the inset,
%we show the details of the TCCSD energies and the total entanglement entropies as
%the orbitals are optimized.
Even at the equilibrium bond length, the wavefunction exhibits a large amount of multi-reference character,
as indicated by the large reduction of roughly 700 mHa in the CASCI energy in the active space alone from the initial HF orbitals to the
final QIO orbitals.
The difference between the CASCI and TCCSD energies can be considered as the dynamic correlation energy,
which is on the order of 1000 mHa across the optimization.

%\section{Original Data}
%\begin{table}{c|c|c|c|c|c}
%Bond lengths $\AA$ &  QIO & TCCSD-NO & CASSCF-NO & UCCSD-NO \\
%1 & 1 & & &
%\end{table}

\onecolumngrid

\section{Original Data} \label{app:data}
See TABLE \ref{tab:c2} for the original data of the C$_2$ molecule and 
TABLE \ref{tab:cr2} for the original data of the Cr$_2$ molecule presented in Figure \ref{fig:c2_cr2_diss} in the 
main text.
\bigskip
%\bigskip
\begin{table}[h]
  \centering
  \begin{tabular}{|l|c|c|c|c|c|c|p{2cm}|p{2cm}|p{2cm}|p{2cm}|p{2cm}|p{2cm}|p{2cm}|}
    \hline
    \rule{0pt}{2.6ex}%\rule[-1.2ex]{0pt}{0pt}
    C-C (Bohr) & QIO  & TCCSD-NO & CASSCF-NO &  UCCSD-NO  & HFO & F$_{\rm QIO} $\\
    \hline
    \rule{0pt}{2.6ex}%\rule[-1.2ex]{0pt}{0pt}
    1.8 & -75.4537845  & -75.4538515 & -75.451787 & -75.4505035 & -75.4477625 & 2.775\\
    2.0 & -75.6339123  & -75.6338969 & -75.631530 & -75.6304366 & -75.6281571 & 2.987\\
    2.2 & -75.7105573  & -75.7102883 & -75.707881 & -75.7067405 & -75.7047199 & 3.190\\
    2.4 & -75.7284669  & -75.7281048 & -75.725810 & -75.7244722 & -75.7226391 & 3.384\\
    2.6 & -75.7151864  & -75.7145051 & -75.712407 & -75.7108455 & -75.7092274 & 3.575\\
    2.8 & -75.6858595  & -75.6858930 & -75.684042 & -75.6823171 & -75.6811697 & 3.634\\
    3.0 & -75.6522752  & -75.6521957 & -75.650593 & -75.6487491 & -75.6485821 & 3.725\\
    3.2 & -75.6342023  & -75.6345553 & -75.632930 & -75.6334080 & -75.6296234 & 2.466\\
    3.4 & -75.6143219  & -75.6144370 & -75.612805 & -75.6131546 & -75.6097061 & 2.607\\
    3.6 & -75.5952101  & -75.5951644 & -75.593422 & -75.5935781 & -75.5904149 & 2.764\\
    3.8 & -75.5779611  & -75.5777775 & -75.575879 & -75.5757977 & -75.5728311 & 2.937\\
    4.0 & -75.5630263  & -75.5628871 & -75.560785 & -75.5604078 & -75.5575338 & 3.133\\
    4.2 & -75.5508078  & -75.5507175 & -75.548390 & -75.5477326 & -75.5447838 & 3.361\\
    4.4 & -75.5412382  & -75.5412656 & -75.538661 & -75.5378033 & -75.5346044 & 3.628\\
    4.6 & -75.5340541  & -75.5341825 & -75.531360 & -75.5304352 & -75.5268035 & 3.922\\
    4.8 & -75.5288546  & -75.5290184 & -75.526062 & -75.5251912 & -75.5210413 & 4.220\\
    5.0 & -75.5250909  & -75.5252870 & -75.522268 & -75.5215623 & -75.5168579 & 4.496\\
    5.2 & -75.5223409  & -75.5225585 & -75.519520 & -75.5190494 & -75.5138422 & 4.732\\
    5.4 & -75.5202896  & -75.5204730 & -75.517464 & -75.5172273 & -75.5116615 & 4.925\\
    5.6 & -75.5186942  & -75.5188673 & -75.515894 & -75.5158398 & -75.5100565 & 5.077\\
    5.8 & -75.5174429  & -75.5175640 & -75.514640  &-75.5147167 & -75.5088626 & 5.196\\
    \hline
  \end{tabular}
  \caption{TCCSD energy and total orbital correlation for C$_2$ presented in Fig. 5 in the main text. Energy in Hartree.}
  \label{tab:c2}
\end{table}

\begin{table}[h]
  \centering
  \begin{tabular}{|l|c|c|c|c|c|p{2cm}|p{2cm}|p{2cm}|p{2cm}|p{2cm}|p{2cm}|}
    \hline
    \rule{0pt}{2.6ex}%\rule[-1.2ex]{0pt}{0pt}
    Cr-Cr ($\rm\AA$) & QIO  & TCCSD-NO & CASSCF-NO &  UCCSD-NO   & F$_{\rm QIO}$\\
    \hline
    \rule{0pt}{2.6ex}%\rule[-1.2ex]{0pt}{0pt}
    1.50 & -2099.8682088 &  -2099.8665733 & -2099.8604731 &  -2099.8505611 & 5.869 \\
    1.55 & -2099.8805779 &  -2099.8790180 & -2099.8715983 &  -2099.8594378 & 6.159 \\
    1.57 & N/A           &  N/A           & -2099.8741577 &  -2099.8609289 & N/A   \\
    1.60 & -2099.8869346 &  -2099.8855955 & -2099.8765241 &  -2099.8614968 & 6.514 \\
    1.64 & -2099.8895896 &  -2099.8882957 & -2099.8777240 &  -2099.8598751 & 6.853 \\
    1.68 & -2099.8907057 &  -2099.8896679 & -2099.8774932 &  -2099.8564156 & 7.226 \\
    1.73 & -2099.8910933 &  -2099.8903517 & -2099.8761989 &  -2099.8506864 & 7.719 \\
    1.80 & -2099.8908079 &  -2099.8903717 & -2099.8735534 &  -2099.8422357 & 8.386 \\
    1.90 & -2099.8897337 &  N/A           & -2099.8690127 &  -2099.8338818 & 9.175 \\
    2.00 & N/A           &  N/A           & -2099.8633707 &  -2099.8345337 & N/A   \\
    \hline
  \end{tabular}
  \caption{TCCSD energy and total orbital correlation for Cr$_2$ presented in Figure \ref{fig:c2_cr2_diss} in the main text. Energy in Hartree.}
  \label{tab:cr2}
\end{table}

\newpage
\twocolumngrid

\bibliography{qicas_2.0}

\end{document}